\documentclass[fleqn,usenatbib]{mnras}

\usepackage{newtxtext,newtxmath}

\usepackage[T1]{fontenc}

\DeclareRobustCommand{\VAN}[3]{#2}
\let\VANthebibliography\thebibliography
\def\thebibliography{\DeclareRobustCommand{\VAN}[3]{##3}\VANthebibliography}

\usepackage{graphicx}	
\usepackage{amsmath}	
\usepackage{subfig}

\title[Magnetic fields in 47 Tuc]{A MeerKAT look at the polarization of 47 Tucanae pulsars: magnetic field implications}

\author[Abbate et al.]{\parbox{\textwidth}{
F.~Abbate,$^{1}$\thanks{E-mail: abbate@mpifr-bonn.mpg.de}
A.~Possenti,$^{2,3}$ 
A.~Ridolfi,$^{2,1}$
V.~Venkatraman~Krishnan,$^{1}$
S.~Buchner,$^{4}$ 
E.~D.~Barr,$^{1}$ 
M.~Bailes,$^{5,6}$
M.~Kramer,$^{1,7}$
A.~Cameron,$^{5,6}$
A.~Parthasarathy,$^{1}$ 
W.~van~Straten,$^{8}$
W.~Chen,$^{1}$
F.~Camilo,$^{4}$
P.~V.~Padmanabh,$^{1,9}$
S.~A.~Mao,$^{1}$
P.~C.~C.~Freire,$^{1}$
S.~M.~Ransom,$^{10}$
L.~Vleeschower,$^{7}$
M.~Geyer,$^{4}$ and
L.~Zhang$^{11,5}$
}
\\
\\
$^{1}$Max Planck Institut f\"ur Radioastronomie, Auf dem H\"ugel 69 D-53121, Bonn, Germany\\
$^{2}$INAF -- Osservatorio Astronomico di Cagliari, Via della Scienza 5,
I-09047 Selargius (CA), Italy\\
$^{3}$Universit\'a di Cagliari, Department of Physics, S.P. Monserrato-Sestu Km 0,700 - I-09042 Monserrato, Italy\\
$^{4}$South African Radio Astronomy Observatory, 2 Fir Street,Cape Town, 7925, South Africa\\
$^{5}$Centre for Astrophysics and Supercomputing, Swinburne University of Technology,\\ 
 Mail H39, PO Box 218, VIC 3122, Australia\\
$^{6}$ARC Center of Excellence for Gravitational Wave Discovery (OzGrav), Swinburne University of Technology, Mail H11, PO Box 218, VIC 3122, Australia\\
$^{7}$Jodrell Bank Centre for Astrophysics, Department of Physics and Astronomy, The University of Manchester, Manchester M13 9PL, UK\\
$^{8}$Institute for Radio Astronomy \& Space Research, Auckland University of Technology, Private Bag 92006, Auckland 1142, New Zealand\\
$^{9}$Max-Planck-Institut f\"{u}r Gravitationsphysik (Albert-Einstein-Institut), D-30167 Hannover, Germany\\
$^{10}$National Radio Astronomy Observatory, 520 Edgemont Rd., Charlottesville, VA, 22903, USA\\
$^{11}$National Astronomical Observatories, Chinese Academy of Sciences,A20 Datun Road, Chaoyang District, Beijing 100101, China\\
}

\date{Accepted XXX. Received YYY; in original form ZZZ}

\pubyear{2022}

\begin{document}
\label{firstpage}
\pagerange{\pageref{firstpage}--\pageref{lastpage}}
\maketitle

\begin{abstract}

  We present the polarization profiles of 22 pulsars in the globular cluster 47 Tucanae using observations from the MeerKAT radio telescope at UHF-band (544-1088 MHz) and report precise values of dispersion measure (DM) and rotation measure (RM). We use these measurements to investigate the presence of turbulence in electron density and magnetic fields. The structure function of DM shows a break at $\sim 30$ arcsec ($\sim 0.6$ pc at the distance of 47 Tucanae) that suggests the presence of turbulence in the gas in the cluster driven by the motion of wind-shedding stars. On the other hand, the structure function of RM does not show evidence of a break. This non-detection could be explained either by the limited number of pulsars or by the effects of the intervening gas in the Galaxy along the line of sight. Future pulsar discoveries in the cluster could help confirm the presence and localise the turbulence.
\end{abstract}




\section{Introduction}

Magnetic fields are found in astronomical objects of all scales. One class of objects where magnetic fields have been very hard to observe and study are globular clusters (GCs). There is very limited evidence of intracluster medium within GCs \citep{Smith1990,vanLoon2006,Barmby2009}. For this reason traditional techniques to detect magnetic fields like synchrotron emission or RM from the gas are not effective. The limited number of studies that have focused on the magnetic fields towards GCs \citep{Abbate2020,Martsen2022} used information derived from pulsars.

Pulsars are rotating neutron stars that emit highly linearly polarised radiation across the radio spectrum. The radio waves from pulsars probe the  intervening matter and its magnetic field between the Earth and the pulsar via two effects: the frequency dependent rotation of the plane of linear polarisation, quantified by the rotation measure (RM) and a frequency dependent dispersion of the radio light, quantified by the dispersion measure (DM). 
The RM is related to the electron density $n_e$ (in cm$^{-3}$) and the line-of-sight component of the magnetic field $B_{\parallel}$ (in $\mu$G) by the equation:
\begin{equation}
{\rm RM}= 0.812 \int_0 ^d n_e(l) B_{\parallel}(l) dl \, \, [{\rm rad \,m^{-2}}],
\end{equation}
where $d$ (in pc) is the distance to the pulsar and $dl$ is the line element along the line of sight. The DM is related to the electron density by the equation:
\begin{equation}
{\rm DM}= \int_0 ^d n_e(l) dl \, \, [{\rm pc \, cm^{-3}}].
\end{equation}
Combining both DM and RM it is possible to estimate the density of free electrons along the line of sight and, under the assumption that magnetic fields and electron densities are not correlated \citep{Seta2021b}, directly probe the component of the magnetic field parallel to the line of sight.

The detection of ionized gas in 47 Tucanae (hereafter 47 Tuc) \citep{Freire2001,Abbate2018} and the very high number of pulsars \citep{Ridolfi2021}\footnote{Currently the second highest pulsar count within GCs with 29 known pulsars. For a complete list of the pulsars in 47 Tuc see: \url{http://www.naic.edu/~pfreire/GCpsr.html} and the references listed there. }, make this GC the ideal target to look for an internal magnetic field. 
The detection of an internal magnetic field would allow us to test which amplification mechanisms are active in these environments. A magnetic field could also influence the searches for radiative signatures of intermediate mass black holes (IMBH) in GCs \citep{Tremou2018}. The magnetic field could affect the accretion rate of the gas on the black hole \citep{Cunningham2012} and, therefore, also the derived mass limits for the IMBH.


A first attempt to observe the magnetic field of 47 Tuc via its pulsars, using observations from the Parkes radio telescope \citep{Abbate2020}, showed the detection of a large gradient in the measured RMs associated with the cluster. The authors attributed it to a possible interaction between the GC and an outflow from the Galactic disk. We refute the presence of the gradient by measuring the RM using significantly more sensitive data from the MeerKAT radio telescope \citep{Jonas2016,Camilo2018}. This cluster has been repeatedly observed by the Large Survey Projects (LSP) MeerTime\footnote{\url{http://www.meertime.org}} \citep{Bailes2016, Bailes2020} and TRansients And PUlsars with MeerKAT (TRAPUM)\footnote{\url{http://www.trapum.org}} \citep{Stappers2016}. A long observational campaign of 23h over three days from 2022 Jan 26 to Jan 29 and an additional observation of 2h on 2022 Jun 9 have been performed on this cluster with the UHF receivers over a frequency range of 544-1088 MHz. The lower central frequency, larger fractional bandwidth and increased sensitivity allowed us to determine RM values for 22 pulsars compared to 13 in the earlier work and with uncertainties on average $\sim 20$ times smaller. 

\section{Observations}

The GC 47 Tuc was the target of an extensive observational campaign from 2022 Jan 26 to Jan 29 with the MeerKAT radio telescope in South Africa with the UHF receivers over a frequency range of 544-1088 MHz. The observations lasted a total of 23 h divided in the following way: 2 h on Jan 26, 17 h on Jan 27, 2 h on Jan 28 and 2 h on Jan 29. An additional observation of 47 Tuc was performed on 2022 Jun 9 lasting 2h. 

\begin{figure*}
    \centering
	\includegraphics[width=0.8\textwidth]{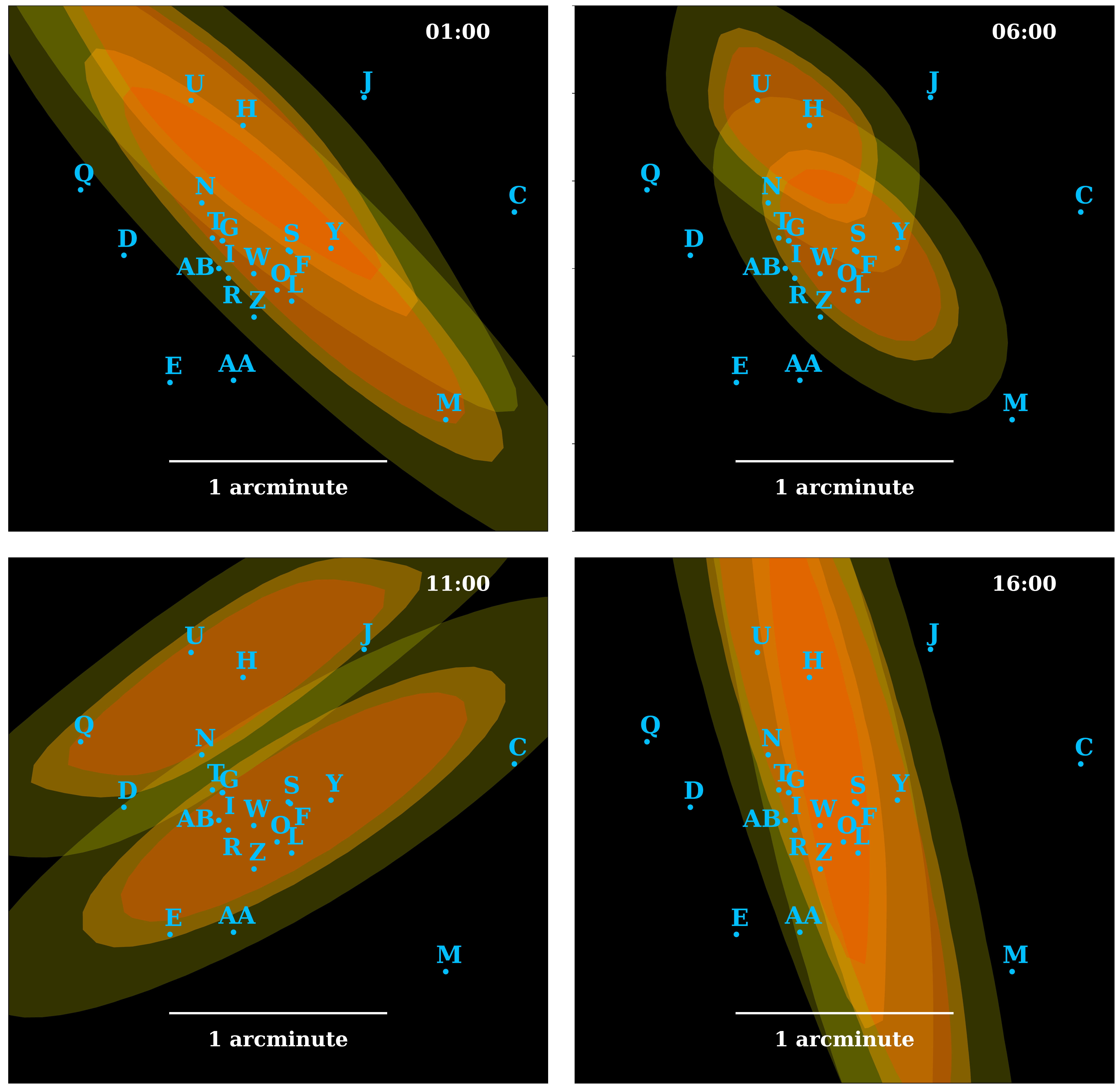}
    \caption{Position and size of the search beams used during the observing campaign of 47 Tuc. The light yellow region shows the half-power level of the beams at 600 MHz, the dark yellow region shows the half-power level of the beams at 900 MHz and the orange region shows the half-power level of the beams at 1100 MHz. The four panels show the evolution of the shape and orientation of the beams during the 17h observation on 2022 Jan 27. The time elapsed from the start of the observation is shown in the top right corner. Some of the pulsars further away from the centre are covered only partially and for short periods of time.}
    \label{fig:beam_pattern}
\end{figure*}

The campaign made use of the Pulsar Timing User Supplied Equipment (PTUSE) machines \citep{Bailes2020} and their ability to synthesize four different tied-array beams to observe different parts of the cluster simultaneously. One beam was pointed at pulsar 47Tuc E and recorded data in timing mode. A second beam, also in timing mode, was pointed at pulsar 47Tuc X for the observation in the first day and then was re-pointed towards pulsar 47Tuc Q. A third beam was pointed towards the centre of the cluster to cover the majority of the pulsars and recorded data in search mode. Finally, the last beam was pointed between pulsars 47Tuc H and 47Tuc U and recorded data in search mode. In the observation on 2022 Jun 9, three beams were set in timing mode pointed at pulsars 47Tuc E, 47Tuc J and 47Tuc AA. The last beam was pointed at the centre of the cluster and recorded data in search mode.
Fig. \ref{fig:beam_pattern} shows the position and shape of the two search beams as they change during the long observation on 2022 Jan 27. 
Some of the pulsars further away from the centre are covered only partially and for short periods of time. In case of weak pulsars or with low linear polarization percentage, like 47Tuc J and 47Tuc AA, this impacted the precision of the RM. In order to get a more precise value, in the observation on 2022 June 9, we pointed a beam in timing mode using all of the available antennas on both of them.

All the beams were initially observed with 4096 frequency channels and coherently de-dispersed at the value reported in Table \ref{tab:setup}. The number of frequency channels of the beams recorded in search mode was reduced on-the-fly by a factor of 16 and only 256 channels were recorded.

\begin{table*}
\centering 
\caption{Setup of the observing campaign on 47 Tuc. For each observing day we report the targets of each beam, the observing mode, the number of antennas used, the sampling time, the recorded number of channels and the value of DM used for coherent de-dispersion. The beam with the target `centre' was pointed at the centre of the cluster while the beam with the target `47 Tuc H,U' was pointed at the halfway point between 47Tuc H and 47Tuc U. }
\label{tab:setup}
\footnotesize
\renewcommand{\arraystretch}{1}
\vskip 0.1cm
\begin{tabular}{cc|cccccc}
\hline
Date   & Beam & Target       &  Mode & \multicolumn{1}{c}{Number of } & \multicolumn{1}{c}{Sampling time} & Number of channels & \multicolumn{1}{c}{DM}\\
 & & & & \multicolumn{1}{c}{antennas} &\multicolumn{1}{c}{($\mu$s)} & & \multicolumn{1}{c}{(pc cm$^{-3}$)}\\
\hline
2022 Jan 26 & 1 & 47Tuc E & Timing & 60 & 7.5 & 4096 & 24.236 \\
& 2 & 47Tuc X & Timing & 60 & 7.5 & 4096 & 24.538 \\
& 3& centre & Search & 50 & 7.5 & 256 & 24.404 \\
& 4& 47Tuc H,U & Search & 52 & 7.5 & 256 & 24.353 \\
\hline
2022 Jan 27 & 1 & 47Tuc E & Timing & 59 & 7.5 & 4096 & 24.236 \\
& 2 & 47Tuc Q & Timing & 59 & 7.5 & 4096 & 24.265 \\
& 3& centre & Search & 49 & 7.5 & 256 & 24.404 \\
& 4& 47Tuc H,U & Search & 51 & 7.5 & 256 & 24.353 \\
\hline
2022 Jan 28 & 1 & 47Tuc E & Timing & 60 & 7.5 & 4096 & 24.236 \\
& 2 & 47Tuc Q & Timing & 60 & 7.5 & 4096 & 24.265 \\
& 3& centre & Search & 50 & 7.5 & 256 & 24.404 \\
& 4& 47Tuc H,U & Search & 52 & 7.5 & 256 & 24.353 \\
\hline
2022 Jan 29 & 1 & 47Tuc E & Timing & 62 & 7.5 & 4096 & 24.236 \\
& 2 & 47Tuc Q & Timing & 62 & 7.5 & 4096 & 24.265 \\
& 3& centre & Search & 52 & 7.5 & 256 & 24.404 \\
& 4& 47Tuc H,U & Search & 54 & 7.5 & 256 & 24.353 \\
\hline
2022 Jun 9 & 1 & 47Tuc E & Timing & 58 & 7.5 & 4096 & 24.236 \\
& 2 & 47Tuc J & Timing & 58 & 7.5 & 4096 & 24.592 \\
& 3& 47Tuc AA & Timing & 58 & 7.5 & 4096 & 24.963 \\
& 4& centre & Search & 41 & 7.5 & 256 & 24.404 \\
\hline
\end{tabular}
\end{table*}

The number of antennas, sampling time, number of channels, and value of DM used for de-dispersion are reported for each beam in Table \ref{tab:setup}. The number of antennas used for the timing beams changed during the campaign based on their availability. For the search beams, the number of antennas was reduced in order to increase the field of view and thus cover a larger area of the cluster. Using simulations of the size and orientation of the beams as they evolved during the observation, we derived an optimal number of antennas to use in order to observe the largest number of pulsars. All of the observations were recorded in full-Stokes so as to recover the polarimetric information.

All the pulsars with the exception of 47Tuc E, 47Tuc Q and 47Tuc X were folded from the data of both beams recorded in search mode using the \texttt{DSPSR}\footnote{{\url{http://dspsr.sourceforge.net}}} pulsar package \citep{vanStraten2011}. For each pulsar we examined the folds in both search-mode beams and kept the one with the highest signal-to-noise ratio (S/N).

The polarization calibration of pulsar observations at MeerKAT is described in \cite{Serylak2021}. This consists of observations of a well known calibrator (either PKS J0408-6545, PKS J0825-5010 or PKS J1939-6342 depending on which one is above the horizon) and of noise diodes in order to obtain the correct delays  and corrections for the two polarizations. During the long observation on 2022 January 27, we repeated the calibration process every 4-5 h in order to maintain a good calibration solution. 
After the observations were taken, we used the routine \texttt{pac} from \texttt{PSRCHIVE}\footnote{\url{http://psrchive.sourceforge.net}} \citep{Hotan2004,vanStraten2012} to correct for the variations of the parallactic angle.
About 8 percent of the channels in the UHF band at MeerKAT are affected by radio frequency interference (RFI). We removed the affected channels using the \texttt{paz} routine of \texttt{PSRCHIVE}.

Using the calibrated and cleaned data, we checked if the ephemerides used for folding were correct. In a few pulsars we could see some drifting of the pulse profile in time. In these cases we extracted times of arrivals (ToAs) using the \texttt{pat} routine of \texttt{PSRCHIVE} and found a local timing solution that removed the drift using \texttt{TEMPO2}\footnote{\url{https://bitbucket.org/psrsoft/tempo2/}} \citep{Hobbs2006}. For the pulsars that showed eclipses during the observations, we removed the orbital phase close to the eclipse since they could impact the linear polarization percentage and the value of RM \citep{Li2022}.

A preliminary measure of the RM of the brightest pulsars using the \texttt{rmfit} routine of \texttt{PSRCHIVE} suggested that the errors on the RM could be as low as 0.1 rad m$^{-2}$. Because of this precision, we were able to see an increase of RM during the day and a decrease during the night by about 0.5-1 rad m$^{-2}$. These are likely caused by the daily variability of the ionosphere \citep{Porayko2019}. The extent of this effect can be predicted using the software \texttt{RMextract}\footnote{\url{https://github.com/lofar-astron/RMextract}} \citep{Mevius2018}. This software estimates the ionospheric RM at a certain position in the sky and time by using a geomagnetic field model, the World Magnetic Model \footnote{\url{https://www.ngdc.noaa.gov/geomag/WMM/DoDWMM.shtml}}, and a global ionospheric map built from Global Navigation Satellite Systems (GNSS) data, the Center for Orbit Determination in Europe global ionospheric map  (CODG) \footnote{\url{https://www.aiub.unibe.ch/research/code___analysis_center/index_eng.html}}.
We divided the data in 20 min chunks in order to keep ionospheric variations in each chunk minimal and estimated the predicted value of the ionospheric RM for each chunk. We applied these corrections to the data using the routine \texttt{pam} of \texttt{PSRCHIVE}. This was enough to remove all the variability in the estimates of RM throughout the observation. Finally, we summed all the observations together using the routine \texttt{psradd} of \texttt{PSRCHIVE} to create a single time-averaged profile of each pulsar with full polarization and frequency information.

We used these profiles to estimate the values of DM for each pulsar. We created one-dimensional analytic templates from the frequency-averaged profiles using the \texttt{paas} routine of \texttt{PSRCHIVE}. We reduced the number of frequency channels to 16 and extracted ToAs for each channel using \texttt{pat}. We performed a fit of DM leaving all other parameters fixed in \texttt{tempo2}.

The RMs were measured in two steps. We first looked at the percentage of linear polarization of the pulsars. An uncorrected RM will cause the polarization position angle to rotate by an angle dependant on the square of the wavelength. This rotation will reduce the degree of linear polarization of the pulsar. We applied different values of RM to the data in order to search for the value that maximizes linear polarization. We searched in the range -100,100 rad m$^{-2}$ with a step size of 0.5 rad m$^{-2}$ to account for the range of values measured in \cite{Abbate2020}. This method has the advantage of being unbiased over a large range of RM and is very effective in the case of pulsars where the linear polarization percentages can be very high. However, the uncertainties are quite large. To find a better estimate of the errors, we corrected the data for this value of RM and we moved on to the second method. We divided the data in 16 frequency channels and performed a fit of the position angle across the frequency band according to the formula: 
\begin{equation}
    \Psi(\lambda) = {\rm RM}\, \lambda^2 + \Psi_0
\end{equation}
where $\Psi_0$ is the value of the position angle as it was emitted by the pulsar. The measure of the value and error of the position angle for each frequency channel follows the prescriptions described in \cite{Noutsos2008,Tiburzi2013,Abbate2020}. In the cases where the position angle changes significantly throughout the pulse profile or shows jumps of 90 deg, we divided the pulse into different regions and performed a simultaneous fit of RM allowing the different regions to have different values of $\Psi_0$. 

As a check of the quality of the fit we also used the routine \texttt{rmfit} to obtain a measure of RM. The different routines returned values compatible within the uncertainties. 

As can be seen in Fig. \ref{fig:beam_pattern}, many of the pulsars were not in the centre of the beam but were observed off-axis. We checked that this does not introduce any bias in the determination of RM by folding a selection of pulsars detected in more than one beam and comparing the RMs. These values are compatible within the uncertainties. We can therefore conclude that the values of RM determined are accurate within the reported uncertainties.

Despite being only 2h long, the additional observation on 2022 Jun 9 was essential to recover accurate RM measurements for 47Tuc J and 47Tuc AA. Both these pulsars are located far from the centre of the cluster and thus only a fraction of their signal was picked up by the search mode beams on the long campaign. The S/N of these pulsars from the observation on 2022 Jun 9, that used two beams in timing mode pointed at these pulsars, was greater than the value obtained from the entire long campaign on 2022 Jan 26-29. The S/N of pulsar 47Tuc J increased by a factor of five while for pulsar 47Tuc AA the improvement was a factor of two. For the other pulsars located at similar distance, like 47Tuc C and 47Tuc M, we were already capable of determining a precise RM from the long campaign.

\section{Results}

Of the 23 pulsars in 47 Tuc with known timing solutions \citep{Pan2016,Ridolfi2016,Freire2017}, 22 pulsars were detected in the observations and for all of them we were able to determine the DM and RM. The polarization profiles of the pulsars are shown in Fig. \ref{fig:pol_profiles}-\ref{fig:pol_profiles_3}. These are obtained correcting each pulsar with the best-fitting RM.

\begin{figure*}
\centering
	\includegraphics[width=\textwidth]{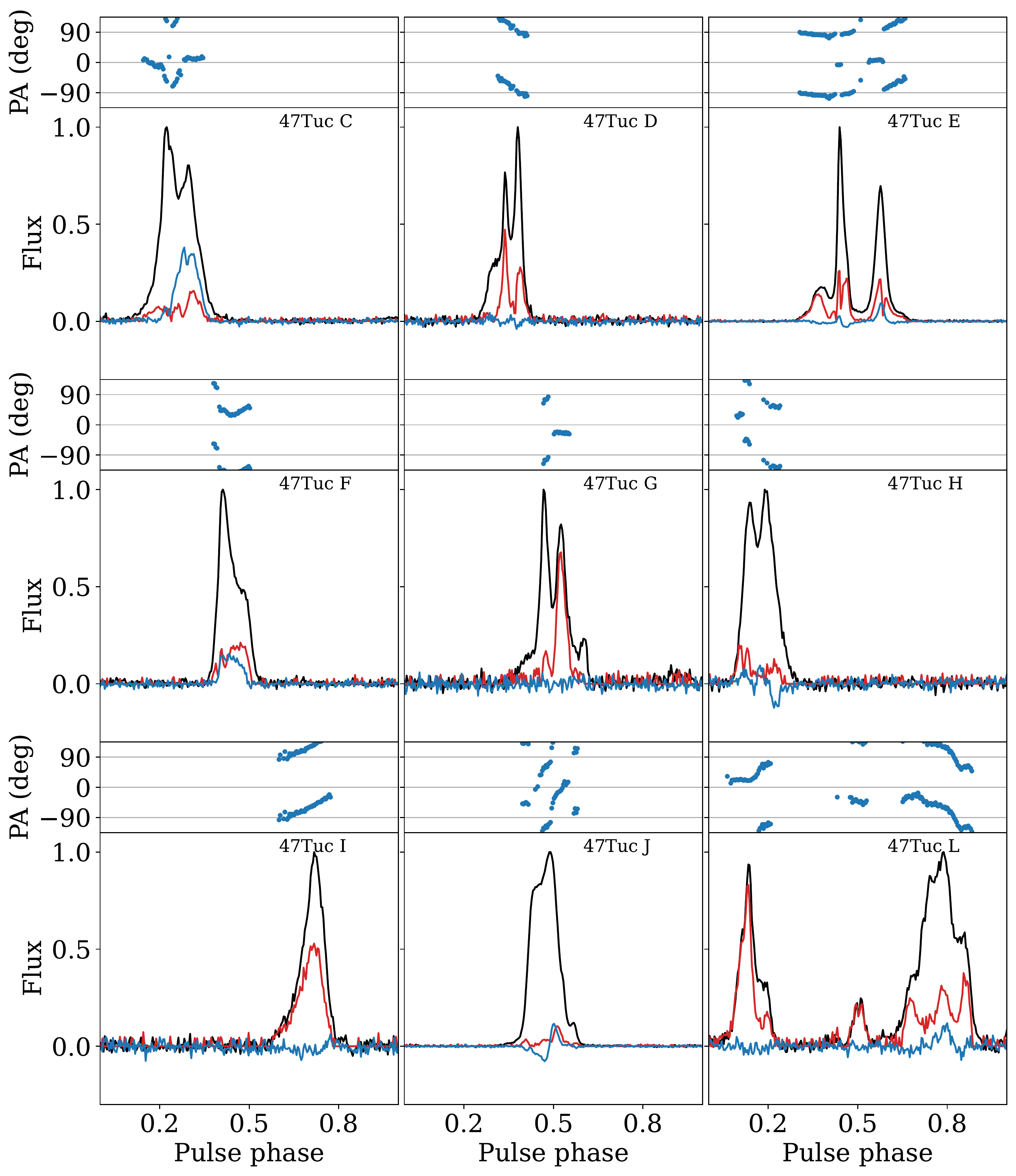}
    \caption{Polarization profiles of the pulsars 47Tuc C - 47Tuc AB. The total intensity (in black) is shown together with the linear polarization (in red) and the circular polarization (in blue). The top panel shows the polarization position angle at infinite frequency and is shown only for the bins where the linear polarization has a S/N higher then 4. The profiles are corrected for the effects of RM.}

  	\label{fig:pol_profiles}
\end{figure*}

\begin{figure*}
    \ContinuedFloat
    \centering
	\includegraphics[width=\textwidth]{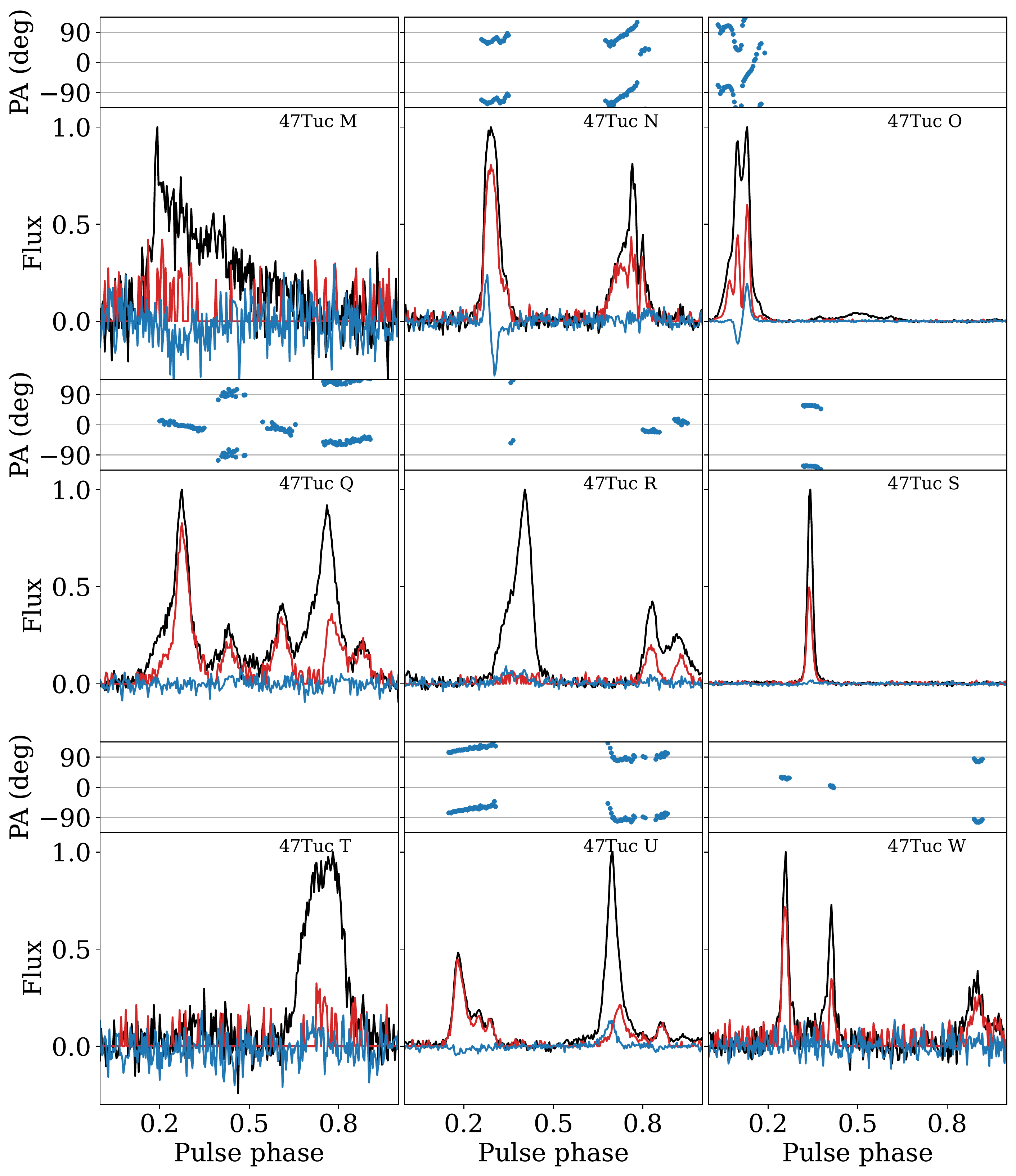}

	\caption{ - continued}
	\label{fig:pol_profiles_2}
\end{figure*}

\begin{figure*}
    \ContinuedFloat
    \centering
	\includegraphics[width=\textwidth]{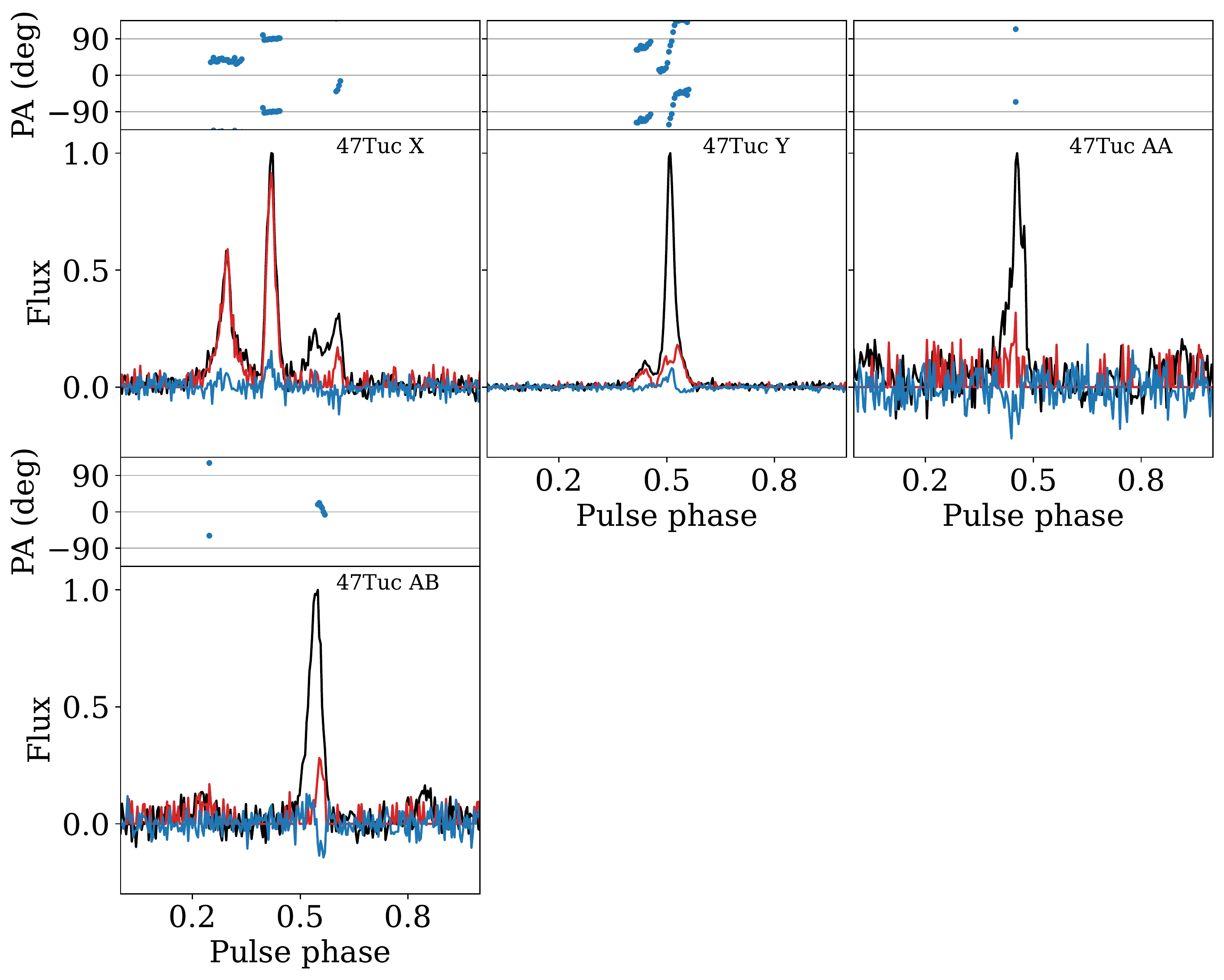}

	\caption{ - continued}
	\label{fig:pol_profiles_3}
\end{figure*}

\begin{figure*}
\centering
	\includegraphics[width=0.8\textwidth]{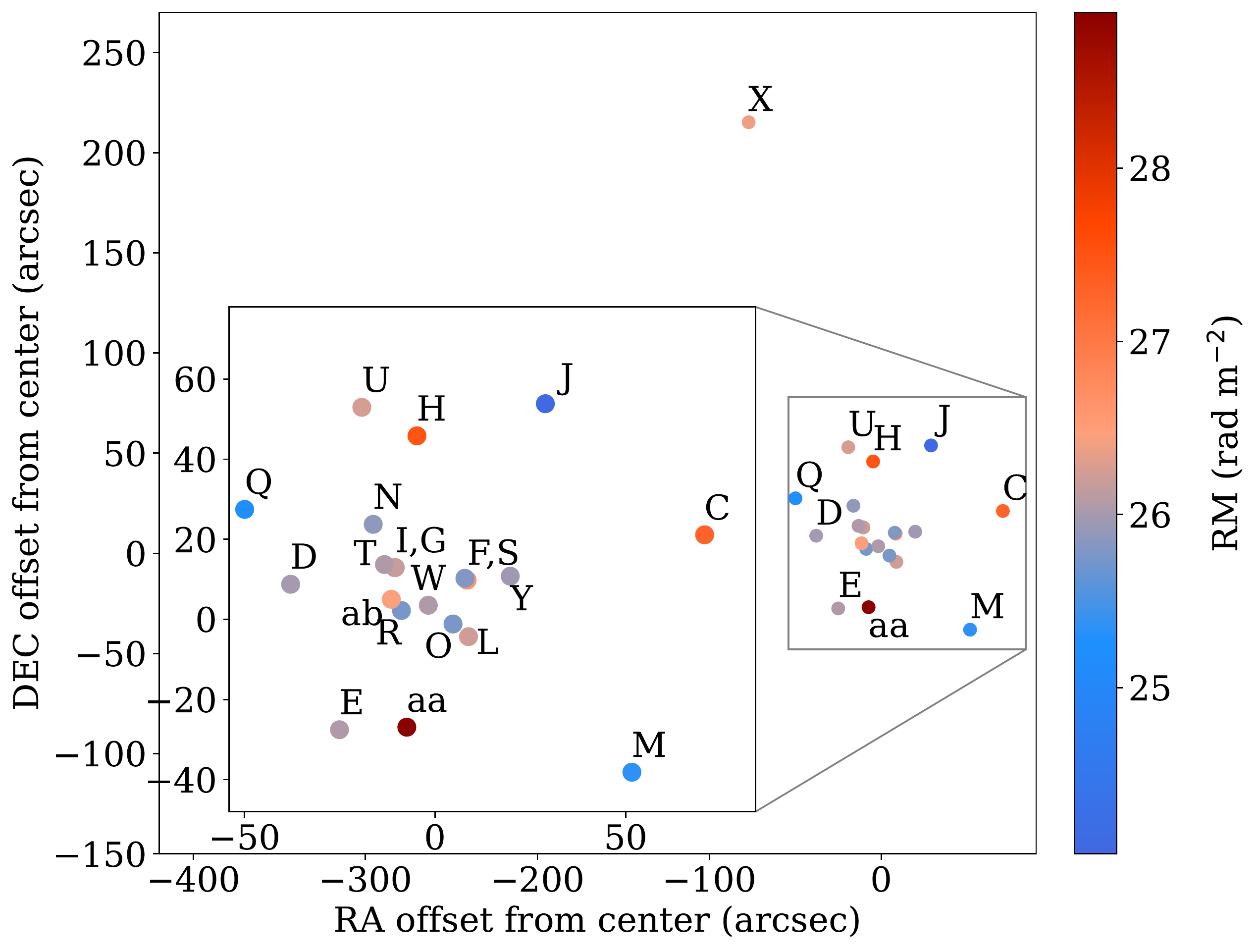}
	\caption{Map of 47 Tuc showing the position of the pulsars with measured RM. The color of the pulsars represents the value of RM according to the color bar at the right of the plot.}
  	\label{fig:RM_map}
\end{figure*}

The results for DM and RM are shown in Table \ref{tab:summary}. A map showing the positions and values of RM for each pulsar is shown in Fig. \ref{fig:RM_map}. The RMs have an average value of $26.2$ rad m$^{-2}$ and a standard deviation of 0.9 rad m$^{-2}$, while the average error is 0.3 rad m$^{-2}$. While the values are not consistent with a single value of RM, the spread is much smaller than previously measured \citep{Abbate2020}. 
To check the reason for this discrepancy, we recovered the original data used in \cite{Abbate2020} and tried to re-measure the RM using the current method. For all pulsars except 47Tuc J, we recovered values compatible with the values from MeerKAT and
not indicative of any gradient. The details of the comparison are reported in the Appendix \ref{comparison}.

\begin{table*}
\caption{DMs and RMs of the pulsars in 47 Tuc detected during the observing campaign. The value in the bracket shows the 1$\sigma$ error on the last digit.}
\label{tab:summary}
\centering
\renewcommand{\arraystretch}{1.0}
\vskip 0.1cm
\begin{tabular}{ccc}
\hline
Pulsar   & \multicolumn{1}{c}{DM}        &  \multicolumn{1}{c}{RM} \\
name        & \multicolumn{1}{c}{(pc cm$^{-3}$)}  & \multicolumn{1}{c}{(rad m$^{-2}$)}\\
\hline
47Tuc C  & 24.5909(3) &  +27.3(2) \\
47Tuc D  & 24.7412(3) &  +26.0(2) \\
47Tuc E  & 24.2396(2) &  +26.0(1) \\
47Tuc F  & 24.3841(3) &  +26.6(2) \\
47Tuc G  & 24.4343(2) &  +25.8(1) \\
47Tuc H  & 24.3750(9) &  +27.5(8)  \\
47Tuc I  & 24.4303(3) &  +26.2(1) \\
47Tuc J  & 24.5937(5) &  +24.0(3) \\
47Tuc L  & 24.3986(6) &  +26.2(1) \\
47Tuc M  & 24.426(3)  &  +25.3(4) \\
47Tuc N  & 24.5568(2)  & +25.9(1) \\
47Tuc O  & 24.3580(1)  & +25.76(9) \\
47Tuc Q  & 24.2794(9)  & +25.2(1) \\
47Tuc R  & 24.36100(8) & +25.7(2) \\
47Tuc S  & 24.38179(1) & +25.80(8) \\
47Tuc T  & 24.421(4)  &  +26.0(7) \\
47Tuc U  & 24.340(1)  &  +26.26(4) \\
47Tuc W  & 24.370(4)  &  +26.0(2) \\
47Tuc X  & 24.538(2)  &  +26.4(2) \\
47Tuc Y  & 24.475(3)  &  +26.0(3) \\
47Tuc AA &  24.921(2)  & +28.9(6) \\
47Tuc AB &  24.3256(5) &  +26.4(6) \\
\hline
\end{tabular}
\end{table*}

\section{Electron density considerations}


The high precision of the values of DM of the pulsars allows us to perform a more detailed analysis of the gas density distribution along the line of sight in the Galactic disk and in the cluster. We check for the presence of turbulence by looking at the second-order structure function (SF) \citep{Minter1996,Haverkorn2008}. The DM SF is defined as:

\begin{equation}
    D_{\rm DM} (\delta \theta)= \langle [{\rm DM}(\vec{u})-{\rm DM}(\vec{v})]^2\rangle,
\end{equation}
where $\vec{u}$ is the position of one source, $\vec{v}$ is the position of a second source with angular separation of $\delta \theta$ and the angle brackets mean the average between all pairs with the same angular separation. In the case of a turbulent medium, we expect the SF to follow either a single power-law or a broken power-law in case the energy injection scale is within the range of observed angular separations \citep{Minter1996,Lazarian2016}.

\begin{figure}
\centering
	\includegraphics[width=0.45\textwidth]{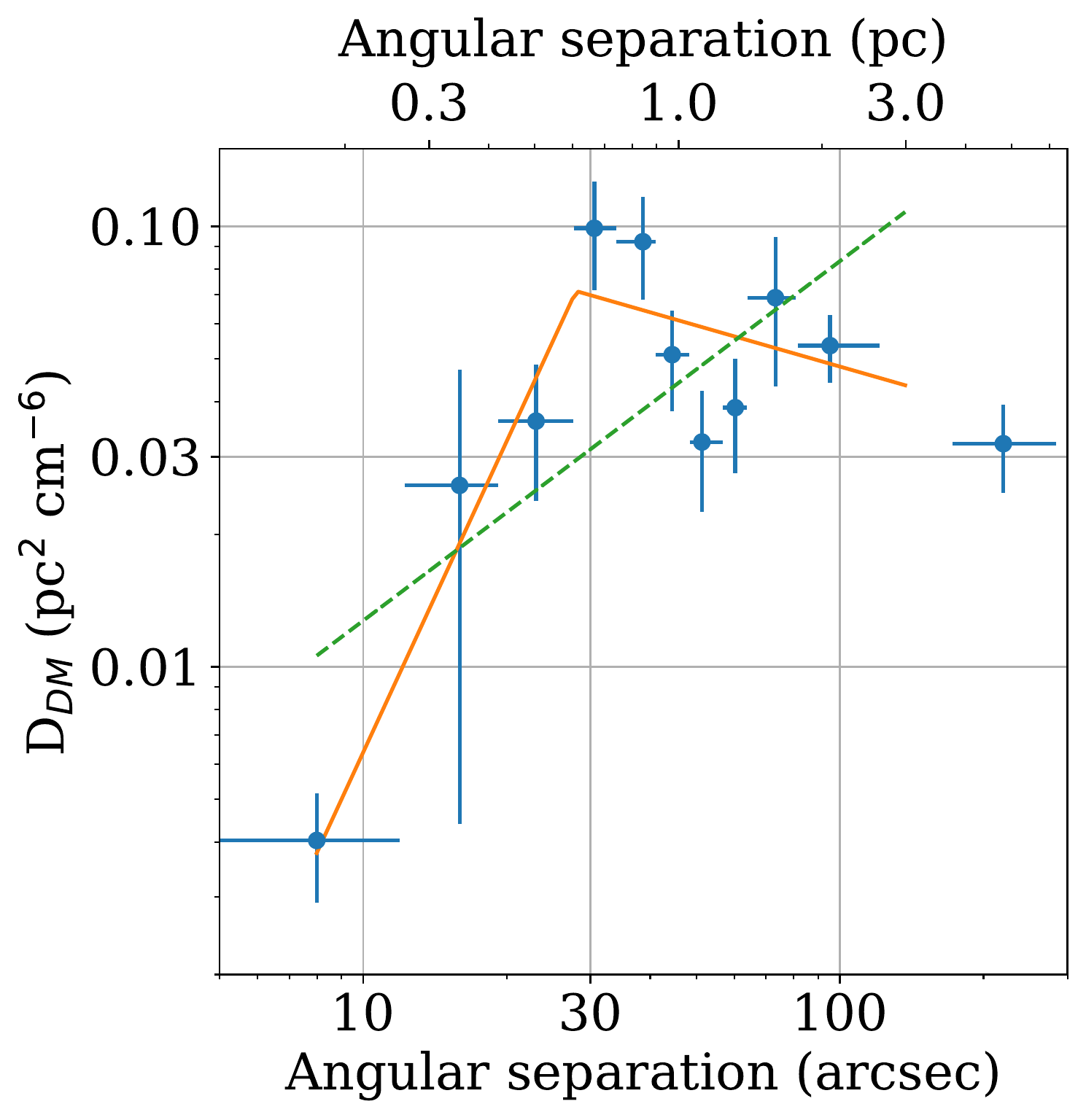}
  	\caption{Plot showing the DM structure function. Each bin represents the average value of 21 DM pairs with similar angular separation. The structure function is fitted with a broken power-law (filled orange line) and a single power-law (dashed green line). The last bin represents the pairs with pulsar 47Tuc X and is excluded from the fits since we cannot assume its value is representative of the average of the DM at this distances.}
  	\label{fig:DM_structure_function}
\end{figure}

To measure the DM SF of the pulsars in 47 Tuc we first estimate the square difference of all DM pairs. We then sort the pairs as a function of angular separation and group the 231 pairs (obtained from the 22 pulsars) together in 10 bins containing each 21 pairs plus 1 bin containing the 21 pairs involving 47Tuc X. Since the angular separations between this pulsar and the cluster centre ($\sim 220$ arcsec) is more than twice the maximum angular separation between any other pair ($\sim 100$ arcsec), we cannot assume its value is representative of the DM at this location. The value and the error of the SF for each bin is measured by taking the mean and the standard deviation of the mean with a Monte Carlo method. We report the DM SF in Fig. \ref{fig:DM_structure_function}.

We further verify the statistical significance of the SF by simulating the effects of the white noise. We repeat the same Monte Carlo extraction as described above assuming the same measured uncertainty and a constant value of DM for each pulsar. The resulting SF has an average value for all of the bins of $3\times 10^{-6}$ rad$^{-2}$ m$^{-4}$, more than three orders of magnitude smaller than the smallest value of the observed DM SF.

We perform a fit to the data with both a single power-law and a broken power-law shown in the plot as the dashed green line and the filled orange line excluding the pairs involving 47Tuc X. 
The fits are based on the Monte Carlo Markov Chain code \texttt{emcee} \citep{Foreman-Mackey2013}. In case of the single power-law fit, the free parameters are the power-law index and the value of the SF at a separation of 1 arcsec. In the case of the broken power-law fit, the free parameters are: the position of the break, the power-law indices before and after the break and the value of the SF at the break. We used a flat prior on the position of the break between the shortest and the largest angular separation, while for the power-law indices we used a flat prior between $-1.5$ and 3.5. We also imposed that the power-law index after the break must be smaller than the value before the break.

In the case of the broken power-law fit, the 68 percent confidence interval of the power-law index before the break is $2.3_{-0.3}^{+0.5}$, for the power-law index after the break it is $-0.3_{-0.3}^{+0.3}$, for the position of the break it is $28_{-7}^{+4}$ arcsec and for the value at the break it is $0.07_{-0.01}^{+0.01}$ pc$^{2}$ cm$^{-6}$. In the case of the single power-law, the 68 percent confidence interval of the power-law index is $0.8_{-0.1}^{+0.1}$ and for the value at 1 arcsec it is $0.002_{-0.001}^{+0.001}$ pc$^{2}$ cm$^{-6}$. We performed an F-test and found a p-value for the single power-law of only 0.02. This value is lower than the threshold of 0.05 so we can reject the hypothesis that the data can be described by a single power-law. 

According to the Kolmogorov theory of turbulence, the presence of a break in the SF could be linked to the energy injection scale of the turbulence \citep{Minter1996,Lazarian2016}. Previous works have suggested that most of the variations in DM between the pulsars are caused by the ionized gas located inside the cluster \citep{Freire2001,Abbate2018}. This suggests that the turbulence is located within the GC itself. At a distance of $\sim 4.5$ kpc \citep{Baumgardt2021}, the angular scale of the break corresponds to a physical size of $0.61_{-0.15}^{+0.09}$ pc.

\subsection{Predictions for the turbulence inside 47 Tuc}

We look in more details at the possible causes and expected values for turbulence in the gas in 47 Tuc. The gas is thought to originate from the winds of evolved Red Giant Branch (RGB) and Asymptotic Giant Branch (AGB) stars \citep{Mcdonald2015}. This material is deposited in the environment of the GC, is ionized and heated by the radiation of young white dwarfs up to $2 \times 10^4$ K and ejected from the cluster in $\sim 4$ Myr \citep{Mcdonald2015,Abbate2018}.

The random motion of wind-shedding stars in the cluster can be a source of turbulent energy in the gas of the cluster \citep{Moss1996}. In this case, the energy injection scale would be the maximum scale at which the gas is deposited in the cluster. We determine it by balancing the ram pressure of the wind and the ambient medium \citep{Baranov1971} using the equation $\rho_w V_w^2 = \rho_{\rm GC} V_{*}^2$, where $\rho_w$ is the density of the wind, $V_w$ is the velocity of the wind, $\rho_{\rm GC}$ is the mass density of the gas in the GC and $V_*$ is the velocity of the RGB stars with respect to the ambient medium. The average velocity of the stars within the half-mass radius, $\sim 6$ pc \citep{Baumgardt2018}, in the cluster is $\sim 20$ km s$^{-1}$, obtained from the one dimensional velocity dispersion \citep{Baumgardt2018} and the radial dependence of velocity dispersion in a King profile (eq. 6 and 7 of \cite{Abbate2018}). The density of the wind depends on the mass loss rate $\dot{M}$, the energy injection scale $R_0$ and the velocity of the wind with the following relation: $\rho_w= \dot{M} /(4 \pi R_0^2 v_w)$. Solving the equation for the energy injection scale we obtain:

\begin{equation}
    R_0= \sqrt{\frac{{\dot{M}} v_w}{4 \pi \rho_{\rm GC} v_*^2}}.
\end{equation}

The mass-loss rate and wind velocity of RGB and AGB stars are dependent on a large number of parameters. Using the modelled values of mass-loss (3-20 $\times 10^{-7} \, {\rm M_{\odot}}$ yr$^{-1}$) and wind velocity (2-4 km s$^{-1}$) for the single brightest RGB and AGB stars in 47 Tuc \citep{McDonald2011b} we obtain values of radii ranging between 0.2-0.5 pc. These values are comparable with the scale of the break observed in the DM SF. 

To estimate the velocity of the turbulence we equate the rate of kinetic energy supplied by the evolved stars to the rate of dissipation of kinetic energy in the turbulence \citep{Moss1996} as follows:
\begin{equation}
    v_t \simeq \left( \frac{R_0 \dot{M}_{\rm tot} v_*^2}{M_{g}}\right)^{1/3},
\end{equation}
where $\dot{M}_{\rm tot} =1.4 \times 10^{-6}$ ${\rm M_{\odot}}$ yr$^{-1}$ is the mass loss from the stars within the half-mass radius of the cluster  \citep{Mcdonald2015} and ${M_{g}} \sim 2$ ${\rm M_{\odot}}$  is the mass of the gas within the half-mass radius \citep{Abbate2018}. 
Using $v_* \sim 20$ km s$^{-1}$, the turbulent velocity becomes $\sim 5.5$ km s$^{-1}$.
The eddy turnover timescale, $\tau \simeq R_0/v_t$, is $\sim 1.1 \times 10^5$ yrs.

\section{Magnetic field considerations}

We now turn our attention to the value of RM of the pulsars in 47 Tuc.
The Galactic contribution to RM estimated from the background sources in the area surrounding 47 Tuc is $+22 \pm 8$ rad m$^{-2}$ \citep{Hutschenreuter2022}. To compare this value with the average value of the pulsars, we first need to estimate how much of the RM seen in background sources originates from gas located in front of the GC. 47 Tuc is located $\sim 4.5$ kpc away \citep{Baumgardt2021} in the Galactic halo at a Galactic latitude of $\sim -45$ deg. The distance from the Galactic disk, $\sim 3.2$ kpc is larger than the scale height of the warm ionized interstellar medium (ISM), 1.6-1.8 kpc \citep{Gaensler2008,Ocker2020} and the scale height of the Galactic magnetic field $2.0 \pm 0.3$ kpc \citep{Sobey2019}. 
Assuming exponentially decreasing disks for the warm ionized ISM and the magnetic field, 91 percent of the entire RM caused by the warm Galactic ISM along this line of sight comes from gas located in front of the cluster. 
The compatibility between the RMs of the pulsars and the value from background sources suggests that the bulk of the RM is caused by the magnetized gas in the Galaxy. We turn our attention to the small but significant differences between the different pulsars.

\subsection{Regular magnetic fields}\label{regular_magnetic_field}

The RM variations between the pulsars could be caused by differences in the number density of electrons and the magnetic field projected along the different lines of sight either in the Galactic disk, in the Galactic halo or in the GC. 

\begin{figure}
    \centering
	\includegraphics[width=0.47\textwidth]{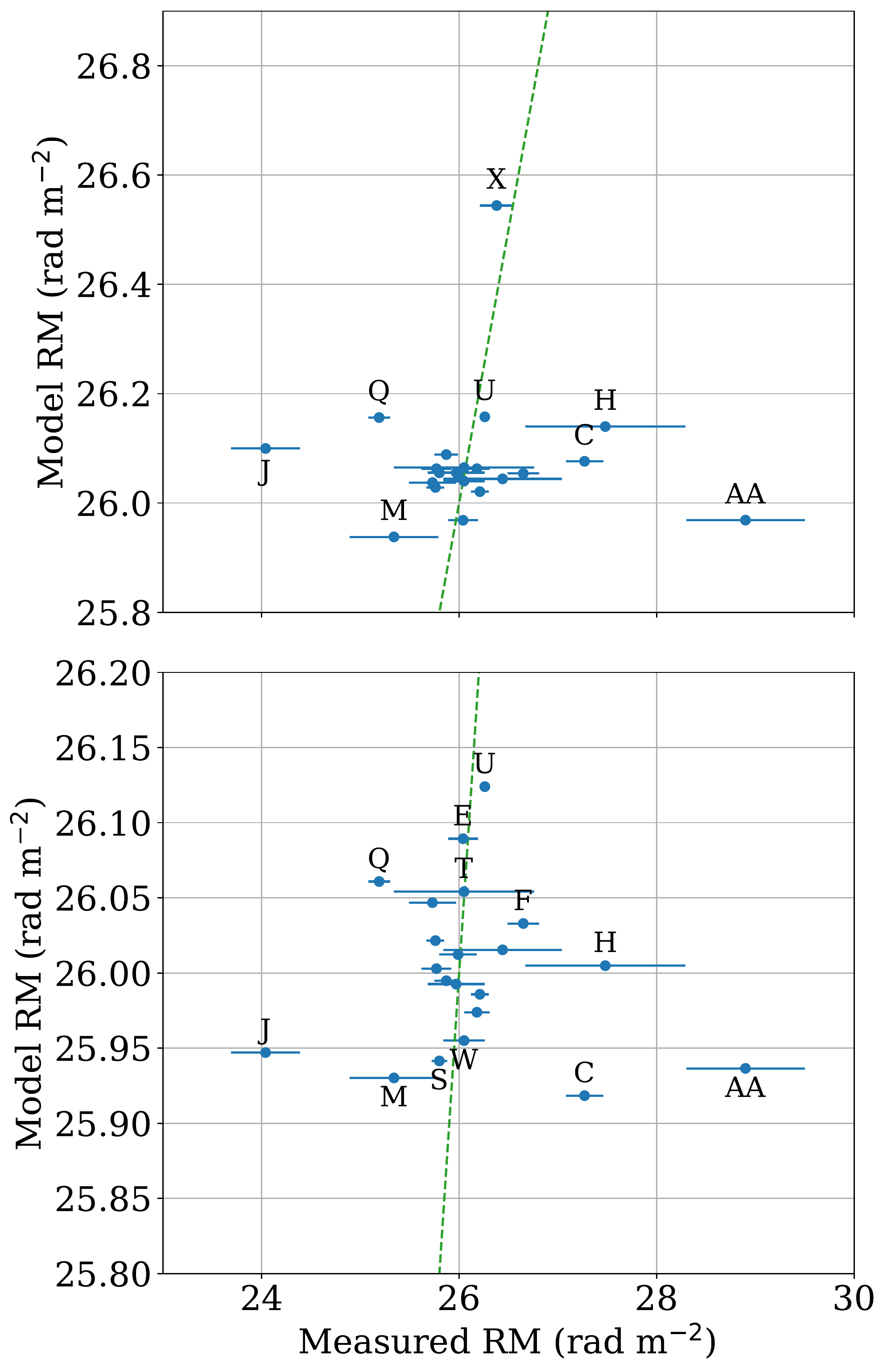}
	\caption{Plots showing the predicted RMs versus the measured RMs in the case of a regular magnetic with different geometries. The top panel shows the results in the case of a linear gradient of RM in the cluster. The bottom panel shows the results in the case of a constant magnetic field inside the cluster. The dashed green line represents the unity line. In the case of a good fit all the points should be compatible with this line. The tested model of regular magnetic field cannot explain the observed RMs.}
	\label{fig:regular_magnetic_fits}
\end{figure}

We first try to see if a regular magnetic field located in the Galactic disk could cause the observed differences. Such a field would look as a linear gradient in RM across the cluster. We perform the fit of RM using as input only the position of the pulsars along the plane of the sky and leaving as free parameters the direction of the gradient, its intensity and the value of RM at the position of the cluster centre. The fit makes use of the Monte Carlo Markov Chain algorithm \texttt{emcee} \citep{Foreman-Mackey2013}. The comparison between the measured RMs and the predictions of the best fit according to this model is shown in the top panel of Fig. \ref{fig:regular_magnetic_fits}. The fit cannot account for most of the RMs of the pulsars. The reduced  $\chi^2$ of the best fit is 12.3.

Another possible source for the RM variations is the magnetic field in the halo. The GC is located at a distance of $\sim 3.2$ kpc below the Galactic disk. At such distances, none of the models of the magnetic field in the halo shows variations at the parsec and sub-parsec scale \citep{Ferrriere2014}. Such a field would contribute a constant quantity to all the pulsars and cannot explain the differences that we see.

Next we look at the case that a magnetic field is located inside the GC. In order for a regular magnetic field to be present in the GC, a large scale dynamo is required \citep{Brandenburg2005}. In 47 Tuc there is evidence of rotation from the stars of $\sim 5$ km s$^{-1}$. However, there is no evidence of differential rotation in the central regions of the cluster \citep{Sollima2019}, so the $\alpha-\Omega$ dynamo is not active. The only dynamo that could be active is the $\alpha^2$ dynamo. This mechanism only requires uniform velocity and its effectiveness is determined by the dynamo number \citep{Brandenburg2005}:
\begin{equation}
    C_{\alpha}=\frac{\alpha R}{\eta_T},
\end{equation}
where $\alpha$ is the mean helicity of the intracluster turbulence, $R$ is the size of the region of interest of the cluster and $\eta_T$ is the turbulent magnetic diffusivity. The mean helicity can be approximated by the following expression \cite{Ruzmaikin1988}: $\alpha \simeq R_0^2 V / R^2$, where $R_0$ is the coherence scale of the turbulence and $V$ is the rotational velocity of the cluster. The turbulent magnetic diffusivity has the form $\eta_T= (1/3) R_0^2/\tau$ \citep{Brandenburg2005}, where $\tau$ is the eddy turnover timescale. The $\alpha^2$ dynamo becomes effective if $C_{\alpha}> C_{\alpha {\rm ,crit}}$, where $C_{\alpha {\rm ,crit}}\simeq 30$ \citep{Brandenburg2005}. Putting everything together we find:
\begin{equation}
    C_{\alpha}\simeq 3 \frac{V\tau}{R}.
\end{equation}

Calculating this factor at the half-mass radius, $R\sim 6$ pc \citep{Baumgardt2018}, we find $C_{\alpha}\sim 0.3$. If instead we restrict ourselves in the region where the pulsars are present, $R \sim 1$ pc, we find $C_{\alpha}\sim 1.6$. In both cases the values are significantly smaller than the critical value.
Therefore, we do not expect this mechanism to be effective in creating and maintaining a large scale magnetic field in 47 Tuc. 

One possible magnetic field configuration that does not require a mean-field dynamo is a magnetic field with constant intensity and direction. This possibility could arise in case the magnetic field in the halo permeates inside the GC. In this case the differences in RM should be caused entirely by differences in the number density of ionized gas. We perform the fit of RM using as input the position of the pulsars along the line of sight measured from the centre of the cluster \citep{Abbate2018} leaving as free parameters the strength of the magnetic field and the value of RM at the centre. The comparison between the measured RMs and the predictions of the best fit according to this model is shown in the bottom panel of Fig. \ref{fig:regular_magnetic_fits}. The model cannot account for the measured RMs with the best fit having a reduced $\chi^2$ of 12. The best fitting value for the strength of magnetic field derived using a constant value of gas density of 0.23 cm$^{-3}$ \citep{Abbate2018} is $-0.6 \pm 0.3$ $\mu$G. This value is compatible with zero at the 2$\sigma$ level but, to be conservative, we take $\sim 1$ $\mu$G as an upper limit on the value of the constant internal magnetic field. 

\subsection{Turbulent magnetic fields}

\begin{figure}
\centering
	\includegraphics[width=0.45\textwidth]{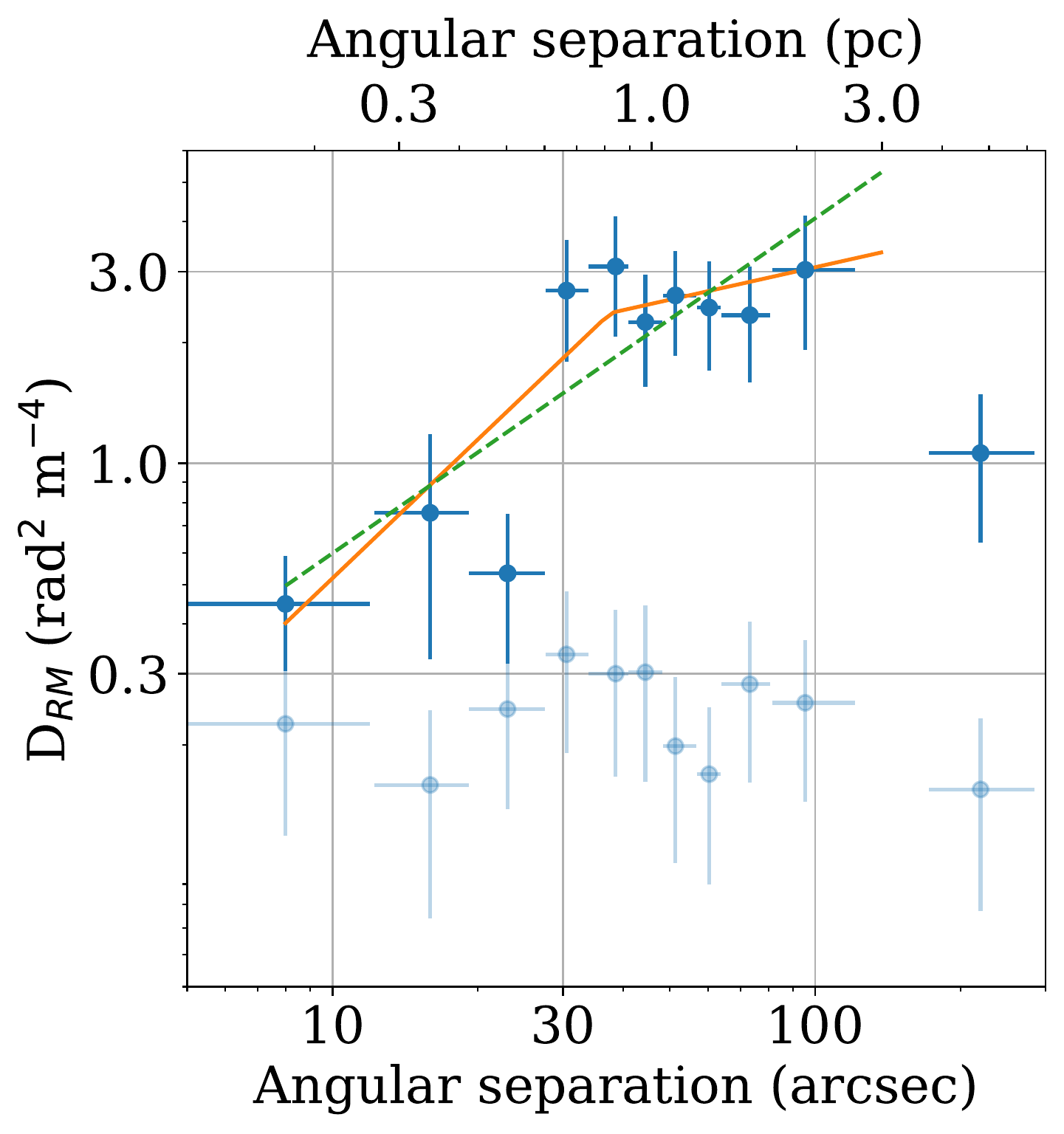}
  	\caption{Plot showing the RM structure function. Each bin represents the average value of 21 RM pairs with similar angular separation. The structure function is fitted with a broken power-law (filled orange line) and a single power-law (dashed green line). The last bin represents the pairs with pulsar 47Tuc X and is excluded from the fits since we cannot assume its value is representative of the average of the RM at this distances. The light blue points show the sensitivity level of the SF determined exclusively by the uncertainties in the measurements. }
  	\label{fig:RM_structure_function}
\end{figure}

In the absence of a regular magnetic field, the observed RM variations could be explained by a turbulent medium. This could be detectable by looking at the RM SF. 
We use the same method and number of bins as the DM case. Also in this case we try fit the data with a broken power-law and a single power-law. The SF with the best fits is shown in Fig. \ref{fig:RM_structure_function}.

Also in this case we repeat the calculation of the sensitivity limit determined by the uncertainties on the measures using the same technique as for DM. For the RMs, the SF of the white noise is of a similar order of magnitude as the observed RM SF and is shown in Fig. \ref{fig:RM_structure_function} as light blue points. For the bins at the smallest angular separation, the sensitivity level is close to the observed value suggesting that the real value could be even lower.

In the case of the broken power-law fit, the 68 percent confidence interval of the power-law index before the break is $1.1_{-0.3}^{+0.5}$, for the power-law index after the break is $0.3_{-0.7}^{+0.4}$, for the position of the break is $38_{-14}^{+66}$ arcsec and the value at the break is  $2.4_{-1.3}^{+0.9}$ rad$^{2}$ m$^{-4}$. In the case of the single power-law, the 68 percent confidence interval of the power-law index is $0.8_{-0.2}^{+0.2}$ and the value at 1 arcsec is $0.09_{-0.04}^{+0.07}$ rad$^{2}$ m$^{-4}$. We performed an F-test and found a p-value for the single power-law of 1 meaning that the extra parameters of the broken power-law fit are not necessary.
Differently from the DM case, there is no detection of a break. This could mean that the observed variations in RM originate from a different environment than the variations of DM or that we are not sensitive enough to detect the break.
In the first case, we check if the variations of RM could be caused by the ISM. In the second case, we estimate the how strong the magnetic field inside 47 Tuc is expected to be and compare it to the observed RMs.

\subsection{Contribution of the ISM to the RM}

The warm ISM is known to show evidence of turbulence \citep{Armstrong1995,Chepurnov2010}.
The RM SF can be described by a turbulent spectrum with an increasing behaviour at small angular scales and almost flat at large angular scales. The transition between the two regimes occurs at the outer turbulence scale. In order to estimate the contribution at the scales observed in the case of 47 Tuc, we need an estimate of the saturation value of the RMSF at large scales, $D_{\rm RM, sat}$, the size of the outer turbulence scale, $r_{\rm out}$, the slope below the break, $\alpha$, and the location of the turbulence in the ISM, $L_T$. 

The the saturation value of the RMSF at large scales can be found from the standard deviation of the RMs from background sources at similar Galactic latitudes \citep{Schnitzeler2010}. For a Galactic latitude of $-45^{\circ}$ the standard deviation of the RMs from the Milky Way is $11.9$ rad m$^{-2}$. The saturation value of the structure function is $D_{\rm RM, sat}\sim2 \sigma_{\rm RM}^2\sim 280$ rad$^{-2}$ m$^{-4}$.

The outer scale of turbulence at mid-Galactic latitudes is expected to be linked to the size of the ejecta from supernova remnants at $r_{\rm out}\sim 90$ pc \citep{Chepurnov2010} with a possible range of $\sim 50-250$ pc \citep{deAvillez2007,Hill2008}. 

The power-law index of the SF before the break depends on the properties of the turbulence. The electron density in the ISM is thought to be well described by a Kolmogorov turbulence with a power-law index of $\alpha=$ 5/3 \citep{Armstrong1995,Chepurnov2010}. However, the RM is caused by both the electron density and the magnetic field, which could follow a different spectrum. Different studies have shown that the electron density fluctuations are dominant with respect to the magnetic field fluctuations \citep{Lazarian2016,Xu2016} so the power-law index is expected to be close to $\alpha=5/3$. In the most general case of turbulence, the lowest value of the power-law index is $\alpha=$1 \citep{Lazarian2016}. This value is larger but still compatible with the best-fitting value of the single power-law fit.  

The turbulence can in principle be located anywhere between us and the GC. Using magnetohydrodynamical simulations, \cite{Hill2008} have found that the observed properties of Galactic H$\alpha$ emission can be recreated by placing the turbulence within a vertical path length of 400-500 pc. At a Galactic latitude of $-45^{\circ}$, this corresponds to $L_T \sim 560-700$ pc.  As an upper limit we choose the scale height of the warm ISM, 1.6-1.8 kpc \citep{Gaensler2008,Ocker2020}. At a Galactic latitude of $-45^{\circ}$, this corresponds to $L_T \sim 2250-2550$ pc.

Putting all of this information together, the contribution of the warm ISM to the SF at the angular scale of $\theta= 30$ arcsec is:
\begin{equation}
    D_{\rm RM,ISM}(\theta)= D_{\rm RM, sat} \left( \frac{\theta L_T}{r_{\rm out}}\right)^{\alpha}.
\end{equation}

Using $D_{\rm RM, sat}= 280$ rad$^{-2}$ m$^{-4}$, $r_{\rm out}= 90$ pc, $\alpha=5/3$ and $L_T= 700$ pc, we find $D_{\rm RM,ISM}(\theta)= 0.005$. This value is only $\sim 0.3$ percent of the value observed. 

Alternatively, we can try to extend the observed single power-law to larger separations until we reach the saturation value. The radius at which the saturation value is reached is $4.5_{-2.7}^{+13.2}$ degrees. Converting it to physical distances using $L_T= 700$ pc, we get an outer scale of $55_{-33}^{+163}$ pc, comparable with the predictions of \cite{deAvillez2007,Hill2008,Chepurnov2010}.  
This suggests that, although the slope is shallower than predicted, the warm ISM could explain the RM variations.



\subsection{Magnetic field inside the GC}

Let's examine the case of a magnetic field internal to the GC. 

If all of the observed variations in RM are caused by a turbulent magnetic field internal to the GC, we can estimate its expected strength. We work under the assumption that the electron density and magnetic field can be described as a constant value plus Gaussian fluctuations with zero mean. For the constant value of the electron density we use a value of $0.23$ cm$^{-3}$ from \cite{Abbate2018} while for the magnetic field we use 1 $\mu$G, the upper limit found for a regular magnetic field in the cluster estimated above. This assumption only works well within a few parsecs of the cluster centre, while, outside, the density rapidly decreases. We estimate the thickness of the region where this assumption would hold, $L$, as the distance at which the density drops to a factor of $e$ of its central value. Using the profile density distribution presented in \cite{Mcdonald2015} we find $L=6$ pc while using the profile of \cite{Abbate2018} we find $L=3$ pc. In the following analysis we consider both values of $L$.

If the RM originates from a region with a thickness of $L$, the variance of RM should be \citep{Xu2016}:
\begin{equation}
    \sigma_{\rm RM}^2 \sim (0.81)^2 (n_e^2 \sigma_B^2 + B_z^2 \sigma_n^2) L^2, 
\end{equation}
where $\sigma_B^2$ and $\sigma_n^2$ are the variances of the turbulent magnetic field and turbulent electron density. With the same assumptions, the variance of electron density can be written as $\sigma_n^2= \sigma_{\rm DM}^2/L^2$. The standard deviation of the turbulent magnetic field becomes 0.8 $\mu$G for $L=6$ pc and 1.6 $\mu$G for $L=3$ pc.

If the evidence of a turbulent medium provided by the DM SF is real, we can estimate the theoretical predictions of the internal magnetic field.
The presence of turbulence in the gas could activate the fluctuation dynamo \citep{Kazantsev1968,Brandenburg2005} and amplify the seed magnetic fields of the stellar winds. This dynamo becomes active in any turbulent medium provided the magnetic Reynolds number is higher than the critical value  $Re_M^{\rm (crit)}\sim 100$ \citep{Federrath2014}. 

We estimate the magnetic Reynolds number $Re_M= v_t R_0 /\eta$ \citep{Moss1996}, where $v_t$ is the velocity of the turbulence, $R_0$ is the energy injection scale and $\eta$ is the Spitzer resistivity that, for a thermal plasma at temperature $T$, is \citep{Brandenburg2005}:
\begin{equation}
    \eta \simeq 10^4 \left( \frac{T}{10^6 {\rm K}}\right)^{-3/2} {\rm cm^2 \, s^{-1}}.
\end{equation}
Using $T = 2 \times 10^4$ K \citep{Mcdonald2015,Abbate2018}, we obtain $\eta \simeq 3.5 \times 10^6 \, {\rm cm^2 \, s^{-1}}$. Using this value in the calculation of the magnetic Reynolds numbers we get $Re_M \sim 3.5 \times 10^{17}$. This is significantly greater than the limit above which the fluctuation dynamo begins to become effective. Fluctuation dynamos are therefore effective at amplifying magnetic fields in 47 Tuc.

A fluctuation dynamo is capable of amplifying magnetic fields by a factor of $e$ every eddy turnover timescale, until the equipartition level between the turbulent energy and the magnetic energy has been reached. This corresponds to a magnetic field of:
\begin{equation}
    B_{\rm eq}=\sqrt{4 \pi \rho v_t^2},
\end{equation}
where $\rho$ is the mass density of the gas. We get  $ B_{\rm eq} \sim 1.4$ $\mu$G. This value is comparable to the estimate of the turbulent magnetic field from the observed RMs and DMs.

After reaching this value, dissipation forces come into play and the amplification stops. The gas stays inside of the cluster only for $\sim 4$ Myr \citep{Mcdonald2015}. This means that the gas goes through $\sim 37$ eddy turnover timescales before being ejected and the seed magnetic field can be amplified up to 16 orders of magnitude. The seed magnetic field of the cluster medium can be found by estimating the magnetic field on the surface of the evolved stars and then applying magnetic flux conservation up to the largest scales reached by the wind. The typical radius of low-mass stars in the RGB phase is $\sim 10 \, {\rm R_{\odot}}$ \citep{Cassisi1997,Mullan2019}. The magnetic field on the surface of these stars is $1-10$ G \citep{Auriere2015,Mullan2019}. Flux conservation would imply that, after reaching a distance of $R_0 \sim 0.6$ pc, the magnetic field would have fallen to $10^{-7}- 10^{-6}$ $\mu$G. In order for this seed field to be amplified up to the equipartition value of $\sim 1$ $\mu$G by the fluctuation dynamo, a time of $1.5-1.7$ Myr is necessary.


Other ways to explain the presence of a magnetic field of a few $\mu$G located in the GC are from the halo magnetic field penetrating in the GC and from local amplification at shock surfaces \citep{Bohdan2021}. However, the thermal pressure from the expanding gas in the GC \citep{Mcdonald2015} prevents the gas from the halo to enter and deposit magnetic fields close to the location of the pulsars. Additionally, the winds from the stars move at speeds (2-4 km s$^{-1}$, \citep{McDonald2011b}) lower than the sound speed ($\sim$ 16 km s$^{-1}$, \citep{Abbate2018}) and the Alfv\'en speed ($v_a = v_t \sim 5.5$ km s$^{-1}$), meaning that there are no shocks where the magnetic fields can be amplified.

\section{Conclusions}

The polarization profiles of the pulsars in 47Tuc obtained with deep MeerKAT observations are shown with unprecedented details.  
The precise measurements of the DM and RM have allowed us to investigate the presence of turbulence in electron density and magnetic fields along the line of sight. 

The DM SF shows evidence of a break at a scale of $\sim 30$ arcsec ($\sim 0.6$ pc at the distance of 47 Tuc). According to the Kolmogorov theory of turbulence \citep{Minter1996, Lazarian2016}, this break could be interpreted as the energy injection scale in a turbulent medium. The turbulence could arise from the stirring of the gas by the random motion of the evolved wind-shedding stars in the GC \citep{Moss1996}. The scale at which these stars deposit gas in the cluster (0.2 -0.5 pc) is comparable with the observed scale of the break.

The RM SF does not show evidence of a break but can be described by a single power-law. This could arise either from the RM contributions of the Galactic ISM or from the gas in the GC. In the hypothesis of the ISM being main contributor, the power-law index of the SF ($\sim 0.8$) is shallower than the expected value of 5/3. However, extending the SF out to the values observed in background sources, the observations are compatible with a value of the outer scale of turbulence in the warm ISM of $\sim 50$ pc, similar to previous predictions \citep{deAvillez2007,Hill2008,Chepurnov2010}.
Alternatively, the variations of the RM could arise from a turbulent magnetic field inside of the GC. The turbulence is strong enough to activate the fluctuation dynamo that could amplify the magnetic fields of the gas up to $\sim 1$ $\mu$G. This value is compatible with the observed variations in RM. In this case the break should be present but we are not sensitive enough to detect it.

The recent pulsar discoveries in the cluster by the MeerTIME/TRAPUM group at MeerKAT \citep{Ridolfi2021}\footnote{The up-to-date list of discoveries can be found at the website: \url{http://www.trapum.org/discoveries/}} could provide a larger number of baselines that can help confirm or rule out the presence of a break in the DM and RM SF. This would allow us to localise the turbulence either in the ISM or in the GC itself.
Turbulent mediums and magnetic fields of similar values should also be present in other GCs but may not be detectable due to the contribution of the ISM that is stronger the closer we move to the Galactic plane. GCs with high Galactic latitudes and with a large number of pulsars like NGC 1851 \citep{Ridolfi2022} and $\omega$ Centauri (Chen et al. in prep) would be the best targets to look for internal gas and magnetic fields.

\section*{Acknowledgements}

The MeerKAT telescope is operated by the South African Radio Astronomy Observatory, which is a facility of the National Research Foundation, an agency of the Department of Science and Innovation. SARAO acknowledges the ongoing advice and calibration of GPS systems by the National Metrology Institute of South Africa (NMISA) and the time space reference systems department department of the Paris Observatory. PTUSE was developed with support from the Australian SKA Office and Swinburne University of Technology. MeerTime data is housed on the OzSTAR supercomputer at Swinburne University of Technology. The OzSTAR program receives funding in part from the Astronomy National Collaborative Research Infrastructure Strategy (NCRIS) allocation provided by the Australian Government. We thank the SARAO technical teams that developed and implemented the 4-beam steerable system used in this work. 
FA, AR, VVK, EDB, MK, APa, WC, PVP, SAM and PCCF acknowledge continuing valuable support from the Max-Planck Society. This work is supported by the Max-Planck Society as part of the "LEGACY" collaboration on low-frequency gravitational wave astronomy.
AR and APo gratefully acknowledge financial support by the research grant ``iPeska'' (P.I. Andrea Possenti) funded under the INAF national call Prin-SKA/CTA approved with the Presidential Decree 70/2016. AR and APo also acknowledge support from the Ministero degli Affari Esteri e della Cooperazione Internazionale - Direzione Generale per la Promozione del Sistema Paese - Progetto di Grande Rilevanza ZA18GR02. The National Radio Astronomy Observatory is a facility of the National Science Foundation operated under cooperative agreement by Associated Universities, Inc. SMR is a CIFAR Fellow and is supported by the NSF Physics Frontiers Center awards 1430284 and 2020265. LZ is supported by ACAMAR Postdoctoral Fellowship and NSFC grant No. 12103069.

\section*{Data Availability}

The data underlying this article will be shared upon reasonable request to the MeerTime and TRAPUM collaborations.



\bibliographystyle{mnras}
\bibliography{biblio} 




\appendix

\section{Comparison with RM results from Parkes reported in Abbate et al. 2020}\label{comparison}

The values of RM reported in this work and the implications for the magnetic field are significantly different from the ones reported in \cite{Abbate2020}. That paper used observations from Parkes in L-band (1208-1520 MHz) using the CASPSR backend to measure the RMs of the pulsars. The authors were able to find RM values for 13 of the pulsars with a total spread over $\sim 40$ rad m$^{-2}$. These data suggested the presence of a gradient in RM across the cluster. The spread in RM for the pulsars in 47 Tuc measured in the present work is only $\sim 5$ rad m$^{-2}$. We searched for the presence of a gradient as discussed in section \ref{regular_magnetic_field}. The resulting best-fitting model in the top panel of Fig. \ref{fig:regular_magnetic_fits} shows no evidence of such gradient.  

RM has a dependence of $f^{-2}$ on the frequency. This means that only considering the different observing bands, 550-1050 MHz for the UHF band at MeerKAT and 1208-1520 MHz for the L-band at Parkes, the uncertainties for MeerKAT would be $\sim 10$ times smaller. Furthermore, the pulsars are brighter at lower frequencies due to their spectral index. This, together with the larger collecting area, implies that the MeerKAT pulsar detections have a significantly higher S/N compared to the Parkes detections, as can be seen comparing the polarization profiles presented in both works. This effect additionally reduces the uncertainty for the RMs measured in the MeerKAT observations.

Only 5 pulsars have RMs incompatible between the two works at the 2$\sigma$ level, namely 47Tuc C, I, J, N and Q. We checked if this discrepancy is real or caused by the technique by which the RMs are measured. We recovered the original calibrated and folded data from Parkes used in \cite{Abbate2020} and redetermined the RM of all pulsars with the technique described above. The comparison between the old results form Parkes, the results from Parkes with the new analysis and the results from the MeerKAT observations are shown in Table \ref{tab:Parkes_RM_results}.
With the only exception of 47Tuc J, we find values compatible at 2$\sigma$ with the ones from MeerKAT observations reported in the current work. We find consistently larger uncertainties than what reported in \cite{Abbate2020}, meaning that their errors might have been underestimated. Furthermore, we note that in their analysis, they did not correct the profiles for the value of RM that maximizes the linear polarization percentage. This might lead to an imprecise determination of the position angle and to a wrong value of RM in the case of low S/N in linear polarization. The new values of RM from the old Parkes observations do not show any indication of a gradient in the direction suggested by \cite{Abbate2020}. 

\begin{table*}
\centering 
\caption{Comparison of the values of RM determined for the pulsars in 47 Tuc from the Parkes observations as reported in \protect\cite{Abbate2020} (first column), from the Parkes observation with the new method (second column) and from the MeerKAT observations presented in this work (third column). The only values not compatible at 2$\sigma$ between the second and third column is 47Tuc J.}
\label{tab:Parkes_RM_results}
\centering
\renewcommand{\arraystretch}{1.0}
\vskip 0.1cm
\begin{tabular}{cccc}
\hline
Pulsar   & \multicolumn{1}{c}{RM reported in \cite{Abbate2020}}  &  \multicolumn{1}{c}{RM re-measured from Parkes} & \multicolumn{1}{c}{RM measured from MeerKAT} \\
name        & \multicolumn{1}{c}{(rad m$^{-2}$)}  & \multicolumn{1}{c}{(rad m$^{-2}$)} & \multicolumn{1}{c}{(rad m$^{-2}$)}\\
\hline
47Tuc C  & +33(2)  & +33(4) &  +27.3(2) \\
\hline
47Tuc D  & +12(12) & +23(9) &  +26.0(2) \\
\hline
47Tuc E  & +27(2)  & +27(9) &  +26.0(1) \\
\hline
47Tuc F  & +18(8)  & +30(8) &  +26.6(2) \\
\hline
47Tuc G  & +12(7)  & +20(13) &  +25.8(1) \\
\hline
47Tuc I  & +5(6)   & +28(10) &  +26.2(1) \\
\hline
47Tuc J  & -9(3)   & -10(6) &  +24.0(3) \\
\hline
47Tuc L  & +19(11) & +26(9) &  +26.2(1) \\
\hline
47Tuc M  & -           & +18(6) &  +25.3(4) \\
\hline
47Tuc N  & +0(6)   & +14(7) & +25.9(1) \\
\hline
47Tuc O  & +24(17) & -      & +25.76(9) \\
\hline
47Tuc Q  & -9(10)   & +37(6)  & +25.2(1) \\
\hline
47Tuc S  & -           & +24(12) & +25.80(8) \\
\hline
47Tuc T  & +12(13) & +13(7) &  +26.0(7) \\
\hline
47Tuc Y  & +24(3)  & +25(8) &  +26.0(3) \\
\hline
\end{tabular}
\end{table*}

The only case where we were not able to recover compatible results is 47Tuc J. The value measured from the Parkes data is $-10 \pm 6$ rad m$^{-2}$ compared to $+24.0 \pm 0.3$ rad m$^{-2}$ measured from the MeerKAT observations presented in this work. In analyzing the different observations made with Parkes, we find that, due to scintillation, most of the S/N comes from two observations: one 2h long on 2014 Aug 20 and one 6h long on 2015 Mar 15. Measuring the RM for these single observations returns different values: $-7 \pm 5$ rad m$^{-2}$ for the observation on 2014 Aug 20 and $+15 \pm 8$ rad m$^{-2}$ for the observation on 2015 Mar 15. We note that the value derived from the second observation is compatible at 2$\sigma$ with the value from the MeerKAT observations. 
This pulsar is in a binary system with orbital period of 2.9h and shows regular eclipses. The observation on 2014 Aug 20 covered the orbital phases around the eclipse while the observation on 2015 Mar 15 covered two entire orbits. The different RMs could be explained by the effects of the material that cause the eclipse.  
We tested this hypothesis by looking at the observation of 47Tuc J taken at MeerKAT on 2022 Jun 9. We divided the observation in 20 min segments and determined the RM for the different segments. The segments far from the eclipsing region showed a value consistent with the orbit-averaged value. On the other hand, the segments close to the eclipse showed lower values of linear polarization. This made it impossible to determine an accurate value of RM for these segments. This effect can cause an inaccurate determination of RM in the Parkes observation of 2014 Aug 20 that occurred around the eclipse. 

We also point to a different reference \citep{Zhang2019} that measured the value of RM for pulsars 47Tuc C, D and J using the Ultra-Wideband Low receiver (704-4032 MHz) at Parkes. For the pulsar 47Tuc J they obtain a value of $20 \pm 4$ rad m$^{-2}$, compatible with the MeerKAT measurement but not with that at the L-band with Parkes. 

We note that a few of the polarization profiles corrected for RM shown in this work present components with almost 100 percent linear polarization. These are 47Tuc G, L, N, Q, U, W and X. Since the ionized gas that causes the RM has the effect of depolarizing the pulsed signal, the true RM corresponds to the maximum polarized signal. We conclude that these values have to be close to the correct value.

\bsp	
\label{lastpage}
\end{document}